\documentclass[12pt]{article}
\pdfoutput=1

\usepackage{jheppub}
\usepackage{graphicx,epsfig,amsmath,amssymb}
\usepackage{subcaption}
\usepackage{enumitem}
\usepackage{calc}
\usepackage[section]{placeins}

\usepackage{array}
\newcommand{\PreserveBackslash}[1]{\let\temp=\\#1\let\\=\temp}
\newcolumntype{C}[1]{>{\PreserveBackslash\centering}p{#1}}
\newcolumntype{R}[1]{>{\PreserveBackslash\raggedleft}p{#1}}
\newcolumntype{L}[1]{>{\PreserveBackslash\raggedright}p{#1}}

\graphicspath{{./plots}}

\def\nn{\nonumber}

\def\be{\begin{equation}}
\def\ee{\end{equation}}
\def\bea{\begin{eqnarray}}
\def\eea{\end{eqnarray}}

\newcommand{\gosam}{\textsc{GoSam}{}}

\newcommand\POWHEGBOX{{\tt POWHEG-BOX}}

\newcommand{\mhh}{m_{hh}}

\newcommand{\ct}{c_t}
\newcommand{\ctt}{c_{tt}}
\newcommand{\chhh}{c_{hhh}}
\newcommand{\cg}{c_{ggh}}
\newcommand{\cgg}{c_{gghh}}
\newcommand{\ftapprox}{FT$_{\mathrm{approx}}$}
\newcommand{\nnloapp}{NNLO$^\prime$} % fix here a notation for the present NNLO
\title{Anomalous couplings in Higgs-boson pair production at approximate NNLO QCD}

\author[a]{Daniel de Florian,}
\author[a]{Ignacio Fabre,}
\author[b]{Gudrun Heinrich,}
\author[c]{Javier Mazzitelli,}
\author[d]{\\Ludovic Scyboz}

\affiliation[a]{International Center for Advanced Studies (ICAS) and
  ICIFI, ECyT-UNSAM, Campus Miguelete, 25 de Mayo y Francia, 1650 Buenos Aires, Argentina}
\affiliation[b]{Institute for Theoretical Physics, Karlsruhe Institute of Technology (KIT), Wolfgang-Gaede-Str. 1, 76131 Karlsruhe, Germany}
\affiliation[c]{Max Planck Institute for Physics, F\"ohringer Ring 6, 80805 M\"unchen, Germany}
\affiliation[d]{Rudolf Peierls Centre for Theoretical Physics, University of Oxford, Parks Road, Oxford OX1 3PU, U.K.}

\emailAdd{deflo@unsam.edu.ar}
\emailAdd{ifabre@unsam.edu.ar}
\emailAdd{gudrun.heinrich@kit.edu}
\emailAdd{jmazzi@mpp.mpg.de}
\emailAdd{ludovic.scyboz@physics.ox.ac.uk}

\preprint{{\small ICAS-61/21, KA-TP-13-2021, P3H-21-48,   MPP-2021-99, OUTP-21-17P}}

\abstract{
  We combine NLO predictions with full top-quark mass dependence with approximate NNLO predictions for Higgs-boson pair production in gluon fusion, including
  the possibility to vary coupling parameters within a non-linear
  Effective Field Theory framework containing five
  anomalous couplings for this process.
We study the impact of the anomalous couplings on various
observables, and present Higgs-pair invariant-mass distributions at seven benchmark points
characterising different $\mhh$ shape types. We also provide numerical coefficients for the 
approximate NNLO cross section as a function of the anomalous couplings at $\sqrt{s}=14$ TeV.
}

\keywords{Higgs couplings, EFT, NNLO}
\begin{document}

\maketitle
\newpage

%\tableofcontents
%\newpage

\section{Introduction}

The Higgs sector is being explored at the LHC with impressive success, having established the coupling of the Higgs boson to all the electroweak gauge bosons \cite{Khachatryan:2016vau}, the top \cite{Sirunyan:2018hoz,Aaboud:2018urx} and bottom quarks~\cite{Aaboud:2018zhk,Sirunyan:2018kst}, the tau lepton~\cite{Sirunyan:2017khh,Aaboud:2018pen} and recently moving towards indication for Higgs couplings to muons~\cite{Sirunyan:2020two,Aad:2020xfq}.
In addition to these important results, it is also crucial to study Higgs boson self-couplings, which provide a way to explore the potential that drives the  electroweak symmetry  breaking mechanism.
The trilinear Higgs boson self-coupling is still rather weakly constrained, even though the limits have improved considerably in Run II~\cite{Sirunyan:2018two,Aad:2019uzh,Sirunyan:2020xok,ATLAS-CONF-2021-016}.
The precision of the measurements of Higgs boson couplings to other particles and itself will increase substantially in the high-luminosity phase of the LHC~\cite{Cepeda:2019klc,DiMicco:2019ngk}. Therefore, simulations of the effects of anomalous couplings in the Higgs sector also need to achieve rather small uncertainties.

Higgs-boson pair production in gluon fusion is a process with direct access to the trilinear Higgs boson self-coupling $\chhh$.
An established deviation from its Standard Model (SM) value would be a clear sign of new physics. However, if $\chhh$ is different from the value predicted by the SM, it is very likely that other Higgs couplings are also modified, such that several anomalous couplings, entering the process $gg\to HH$ dominantly within an Effective Field Theory (EFT) framework, should be studied simultaneously.
Furthermore, in order to get reliable predictions, higher-order QCD corrections need to be taken into account, not only for the SM cross section but also in the case anomalous couplings are included.

The loop-induced process $gg\to HH$ at leading order has been calculated in Refs.~\cite{Eboli:1987dy,Glover:1987nx,Plehn:1996wb}. 
Before the full next-to-leading order (NLO) QCD corrections became available, the $m_t\to\infty$ limit (``Heavy Top Limit, HTL''), sometimes also called ``Higgs Effective Field Theory~({\sc heft})'' approximation, has been used. 
In this limit, the NLO corrections were first calculated in 
Ref.~\cite{Dawson:1998py} using the so-called ``Born-improved HTL'', 
which involves rescaling the NLO results in the $m_t\to\infty$ limit by the LO result in the full theory.
In Ref.~\cite{Maltoni:2014eza} an approximation called
``\ftapprox'' was introduced, which contains the real radiation matrix elements 
with full top-quark mass dependence, while the virtual part is
calculated in the Born-improved HTL approximation.
The NLO QCD corrections with full top-quark mass dependence became available more recently~\cite{Borowka:2016ehy,Borowka:2016ypz,Baglio:2018lrj,Baglio:2020ini}.
The NLO results of Refs.~\cite{Borowka:2016ehy,Borowka:2016ypz} have been combined with parton shower Monte Carlo programs in Refs.~\cite{Heinrich:2017kxx,Jones:2017giv,Heinrich:2019bkc,Heinrich:2020ckp}, where Ref.~\cite{Heinrich:2020ckp} introduced the possibility of varying five Higgs couplings.

In the $m_t\to\infty$ limit, the next-to-next-to-leading order (NNLO) QCD corrections have been computed in Refs.~\cite{deFlorian:2013uza,deFlorian:2013jea,Grigo:2014jma,Grigo:2015dia,deFlorian:2016uhr}.
 The calculation of Ref.~\cite{deFlorian:2016uhr} has been combined with results including the top-quark mass dependence as far as available in Ref.~\cite{Grazzini:2018bsd}, defining an NNLO$_{\mathrm{FTapprox}}$ result which contains the full top-quark mass dependence at NLO as well as in the double real radiation part.
 Soft gluon resummation combined with these results has been presented in Ref.~\cite{deFlorian:2018tah}.
 N$^3$LO corrections are also available~\cite{Chen:2019lzz,Chen:2019fhs}, where in Ref.~\cite{Chen:2019fhs} the N$^3$LO results in the HTL have been ``NLO-improved'' using the results of Refs.~\cite{Heinrich:2017kxx,Heinrich:2019bkc}.
 In Ref.~\cite{deFlorian:2017qfk},  NNLO results in the Born-improved heavy top limit including the effect of anomalous couplings were presented, see also Ref.~\cite{Grober:2015cwa} for the NLO case.

The scale uncertainties at NLO are still at the 10\% level, while they are decreased to about 5\% when including the NNLO corrections and to 
about 3\% at N$^3$LO in the ``NLO-improved'' variant.
The uncertainties due to the chosen top mass scheme have been assessed in Refs.~\cite{Baglio:2018lrj,Baglio:2020ini,Baglio:2020wgt}.

In this work we present a study of the anomalous couplings relevant to the process $gg\to HH$ within a non-linear EFT operator expansion, at approximate NNLO in QCD (which we will denote by \nnloapp{}).
It builds on the results presented in  Ref.~\cite{deFlorian:2017qfk}, and extends them to include the full NLO QCD corrections from Ref.~\cite{Buchalla:2018yce}. 
Thus \nnloapp{} contains NNLO results in the heavy top limit, where exact Born expressions have been used whenever the higher-order corrections in the HTL factorise, as well as NLO corrections with full top-quark mass dependence.

In particular, we provide coefficients for all the possible combinations of anomalous couplings that can occur at NNLO for the total cross section, analogous to what has been provided at NLO in Ref.~\cite{Buchalla:2018yce}.
These coefficients can then be used to reconstruct the cross section for any combination of coupling values.
Furthermore, we show results for seven benchmark points characteristic of certain shape types of the Higgs-boson pair invariant-mass distribution $\mhh$, which have been identified by an NLO shape analysis presented in Ref.~\cite{Capozi:2019xsi}.

This paper is organised as follows. In Section~\ref{sec:couplings} we  describe the theoretical framework and the definition of the anomalous couplings. Section~\ref{sec:results} contains the phenomenological results, including total cross sections and differential distributions for the benchmark points, and the description of the fitting procedure for the coefficients of the anomalous couplings, together with the corresponding results. Finally, we conclude in Section~\ref{sec:conclusions}.

\section{Higgs-boson pair production within an EFT framework}
\label{sec:couplings}

The calculation builds on the ones presented in
Refs.~\cite{Buchalla:2018yce,Heinrich:2020ckp,deFlorian:2017qfk} and
therefore the methods will be described only briefly here, focusing on
the new aspects.

\begin{figure}[h]
\begin{center}
\includegraphics[width=11cm]{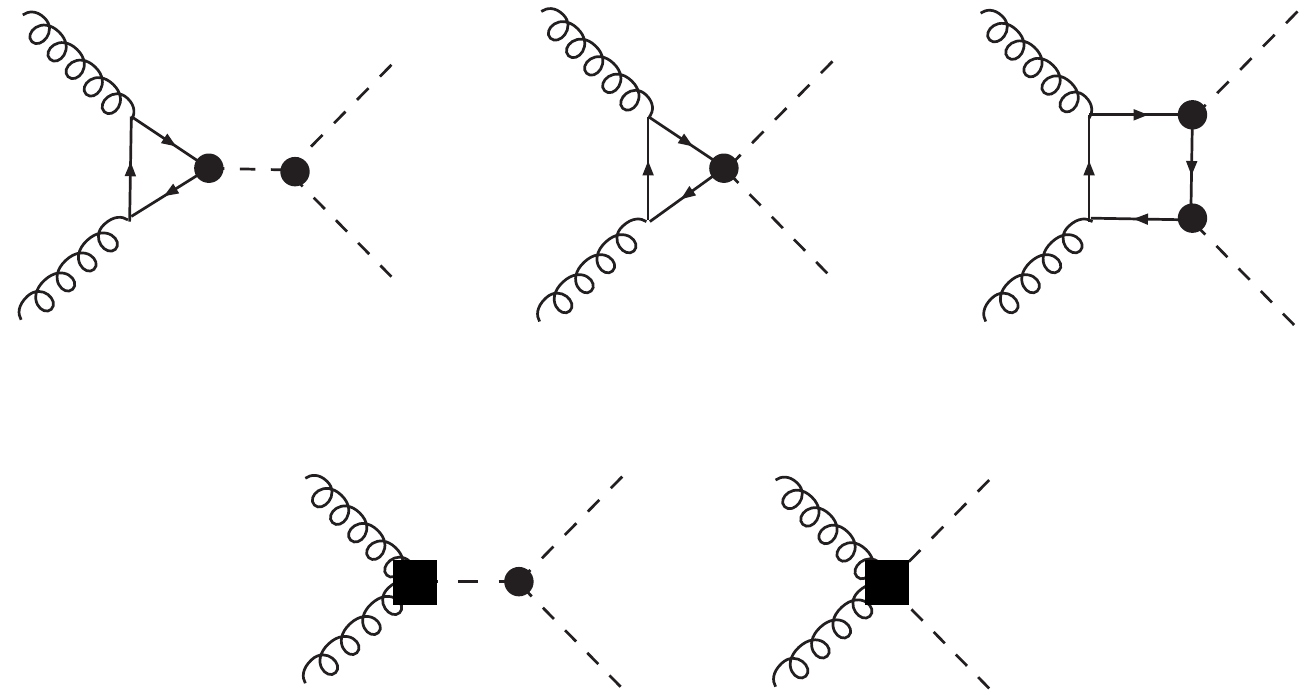}
\end{center}
\caption{Higgs-boson pair production in gluon fusion at leading order in QCD and at chiral dimension $d_\chi=4$. The black dots indicate vertices from
anomalous couplings present already at leading order in the chiral Lagrangian ($d_\chi=2$),
the black squares denote local operators contributing at ($d_\chi=4$).}
\label{fig:hprocess}
\end{figure}

We work in a non-linear EFT framework, sometimes also called
Electroweak Chiral Lagrangian (EWChL)
including a light Higgs boson~\cite{Alonso:2012px,Buchalla:2013rka} or
HEFT (Higgs Effective Field Theory), not to be confused with the heavy top limit, which is sometimes also called {\sc heft}.
It relies on counting the chiral dimension of the terms
contributing to the Lagrangian~\cite{Buchalla:2013eza}, rather than counting the canonical
dimension as in the Standard Model Effective Field Theory (SMEFT). As a consequence, the EWChL is also suitable to describe strong dynamics in the Higgs sector.
Applying this framework to Higgs-boson pair production in gluon
fusion, keeping terms up to, and including, chiral dimension $d_\chi = 4$, we obtain  the
effective Lagrangian relevant to this process as
\begin{align}
{\cal L}\supset 
-m_t\left(c_t\frac{h}{v}+c_{tt}\frac{h^2}{v^2}\right)\,\bar{t}\,t -
c_{hhh} \frac{m_h^2}{2v} h^3+\frac{\alpha_s}{8\pi} \left( c_{ggh} \frac{h}{v}+
c_{gghh}\frac{h^2}{v^2}  \right)\, G^a_{\mu \nu} G^{a,\mu \nu}\;.
\label{eq:ewchl}
\end{align}
In the EWChL framework there are a priori no relations between the
couplings. In general, all couplings may have arbitrary values of ${\cal O}(1)$.
The conventions are such that in the SM $c_t=c_{hhh}=1$ and $c_{tt}=c_{ggh}=c_{gghh}=0$. The EWChL coefficients can be related~\cite{Grober:2015cwa} to those in the SMEFT at Lagrangian level,
however how to treat double insertions of operators and squared dimension-6 terms at cross section level is less straightforward when attempting to relate the two EFT frameworks.
The diagrams at leading order in QCD and chiral dimension four are shown in Fig.~\ref{fig:hprocess}.

There are different normalisation conventions for the anomalous couplings in the literature.
In Table~\ref{tab:conventions} we summarise some conventions commonly used.
\begin{table}[htb]
\renewcommand{\arraystretch}{1.1}
\begin{center}
\begin{tabular}{ |c | c |c| }
\hline
Eq.~(\ref{eq:ewchl}), i.e. ${\cal L}$ of Ref.~\cite{Buchalla:2018yce}& Ref.~\cite{Carvalho:2015ttv}&Ref.~\cite{Grober:2015cwa}\\
\hline
$c_{hhh}$ & $\kappa_{\lambda}$ & $c_3$\\
\hline
$c_t$ &$ \kappa_t$  &$c_t$\\
\hline
$ c_{tt} $ & $ c_{2}$  &$c_{tt}/2$\\
\hline
$c_{ggh}$ &$ \frac{2}{3}c_g $  &$8c_{g}$\\
\hline
$c_{gghh}$ & $-\frac{1}{3}c_{2g}$ &$4c_{gg}$\\
\hline
\end{tabular}
\end{center}
\caption{Translation between different conventions for the definition of the anomalous couplings.\label{tab:conventions}}
\end{table}
For the relation to the corresponding parameters in the SMEFT we refer
to Ref.~\cite{Heinrich:2020ckp}.

In Ref.~\cite{Buchalla:2018yce}  the NLO QCD corrections were calculated within this framework.
These corrections have been implemented into the code {\tt
  ggHH}~\cite{Heinrich:2019bkc,Heinrich:2020ckp}, which is a public code available at~\cite{powhegurl}, where 
 the real radiation matrix elements were implemented
using the interface between \gosam~\cite{Cullen:2011ac,Cullen:2014yla} and
the \POWHEGBOX~\cite{Frixione:2007vw,Alioli:2010xd,Luisoni:2013cuh} and the virtual two-loop corrections use the results of the
calculations presented in Refs.~\cite{Borowka:2016ehy,Borowka:2016ypz}.
In turn, approximate NNLO QCD corrections in a similar EFT framework have been computed in Ref.~\cite{deFlorian:2017qfk}, working in the heavy top limit improved by inserting LO form factors with full top mass dependence.
The results described in this paper are obtained by performing an additive combination of the ones presented in Refs.~\cite{Buchalla:2018yce,deFlorian:2017qfk}, i.e.~keeping the full top mass dependence up to NLO and {\it adding} the approximate results for the genuine NNLO piece.
We will denote our results by \nnloapp.

We can describe the dependence of the NNLO cross section on the five anomalous couplings in terms of 25 coefficients $a_i$, following Refs.~\cite{Azatov:2015oxa,Carvalho:2015ttv,Buchalla:2018yce}:
\begin{align}
\sigma_{\rm BSM}/\sigma_{\rm SM}  &= 
a_1\, c_t^4 + a_2 \, c_{tt}^2  + a_3\,  c_t^2 \chhh^2  +
a_4 \, \cg^2 \chhh^2  + a_5\,  \cgg^2  +
a_6\, c_{tt} c_t^2 + a_7\,  c_t^3 \chhh \nn\\
& + a_8\,  c_{tt} c_t\, \chhh  + a_9\, c_{tt} \cg \chhh + a_{10}\, c_{tt} \cgg +
a_{11}\,  c_t^2 \cg \chhh + a_{12}\, c_t^2 \cgg \nn\\
& + a_{13}\, c_t \chhh^2 \cg  + a_{14}\, c_t \chhh \cgg +
a_{15}\, \cg \chhh \cgg\, + a_{16}\, c^3_t \cg \nn\\
& + a_{17}\,  c_t c_{tt} \cg
+ a_{18}\, c_t \cg^2 \chhh + a_{19}\, c_t \cg \cgg \, + a_{20}\,  c_t^2 \cg^2
\nn\\
&+ a_{21}\, c_{tt} \cg^2
        + a_{22}\, \cg^3 \chhh + a_{23}\, \cg^2 \cgg
        + a_{24}\, \cg^4 + a_{25}\, \cg^3 \ct  \,.
\label{eq:AcoeffsNNLO}
\end{align}
While at LO only the first 15 coefficients contribute, at NLO 23
coupling combinations occur and at NNLO 25 combinations are possible.

It is worth stressing that the functional dependence described in Eq.~(\ref{eq:AcoeffsNNLO}), which holds for the exact NNLO cross section, is also valid for the approximate results presented in this work. To this end, the fact that the combination of the full NLO and approximate NNLO is done through an additive approach becomes crucial, since e.g. a simple multiplicative rescaling would not respect the functional form of Eq.~(\ref{eq:AcoeffsNNLO})
(and, in particular, due to the new combinations appearing at NNLO it would be ill-defined for certain values of the anomalous couplings).

\section{Phenomenological results and coupling coefficients}
\label{sec:results}

Our results are calculated at a centre-of-mass energy of
$\sqrt{s}=14$\,TeV. The parton distribution functions
PDF4LHC15~\cite{Butterworth:2015oua,CT14,MMHT14,NNPDF}
 have been used, interfaced via
LHAPDF~\cite{Buckley:2014ana}, along with the corresponding value for
$\alpha_s(\mu)$.
We use NLO parton densities and strong coupling evolution for the LO and NLO predictions, and NNLO PDFs and strong coupling evolution for the \nnloapp{} calculation.
The masses of the Higgs boson and the top quark have been
fixed to $m_h=125$\,GeV, $m_t=173$\,GeV and their widths have been set to zero.
The top-quark mass is renormalised in the on-shell scheme.

The scale uncertainties are
estimated by varying the factorisation/renormalisation scales
$\mu_{F}, \mu_{R}$, where the bands 
represent 3-point scale variations around the central scale $\mu_0 =\mhh/2$, with
$\mu_{R} = \mu_{F}=\xi\,\mu_0$, where $\xi \in \{0.5,1,2\}$.

\subsection{Total cross section}

In Ref.~\cite{Capozi:2019xsi}, shapes of the Higgs-boson pair invariant-mass distribution $\mhh$ were ana\-lysed in the 5-dimensional
coupling parameter space,  using machine learning techniques to
classify $\mhh$-shapes from NLO predictions.
This procedure led to seven shape characteristic benchmark points, for which we also
show \nnloapp{}  results. For convenience we repeat the
benchmark points in Table~\ref{tab:benchmarks}, together with the
corresponding values for the NLO and \nnloapp{} cross sections at  $\sqrt{s}=14$\,TeV.

\begin{table}
\renewcommand{\arraystretch}{1.5}
\resizebox{\textwidth}{!}{  
\begin{tabular}{|c|C{0.9cm}|C{0.9cm}|C{0.9cm}|C{0.9cm}|C{0.9cm}|c|c|c|c|c|}
\hline
\!{\tiny benchmark}\! & $c_{t}$ & $c_{hhh}$ & $c_{tt}$ & $c_{ggh}$ & $c_{gghh}$ &
                                                                      $\sigma_{\rm{NLO}}$ [fb] &  $\sigma_{\rm{NNLO}^\prime}$ [fb] & $K_{\rm{NLO}}$ & $K_{\rm{NNLO}^\prime}$ & ratio to SM \\
\hline
SM & 1 &  1 &  0 & 0 &  0 & $32.90^{+14\%}_{-16\%}$ & $36.69^{+0.0\%}_{-4.3\%}$ & 1.66 & 1.85 & 1.00 \\
\hline
1 & 0.94 &  3.94 &  -$\frac{1}{3}$ & 0.5 &  $\frac{1}{3}$ & $222.6^{+18\%}_{-14\%}$  & $237.2^{+2.7\%}_{-5.4\%}$ & 1.90 & 2.03 & 6.47 \\
\hline
2 & 0.61 & 6.84 & $\frac{1}{3}$ &  0.0 & -$\frac{1}{3}$ & $168.1^{+20\%}_{-16\%}$ & $191.1^{+7.1\%}_{-8.6\%}$ & 2.14 & 2.43 & 5.21 \\
\hline
3 & 1.05 &  2.21 &  -$\frac{1}{3}$ & 0.5 & 0.5 & $151.9^{+17\%}_{-14\%}$  & $159.9^{+2.1\%}_{-5.2\%}$ & 1.84 & 1.92 & 4.36 \\
\hline
4 & 0.61 &  2.79 &  $\frac{1}{3}$ &  -0.5 & $\frac{1}{6}$ & $63.14^{+20\%}_{-16\%}$  & $69.57^{+8.9\%}_{-9.1\%}$ & 2.14 & 2.37 & 1.90 \\
\hline
5 & 1.17 &  3.95 &  -$\frac{1}{3}$ & $\frac{1}{6}$ &  -0.5 & $154.8^{+14\%}_{-13\%}$  & $166.7^{+0.0\%}_{-3.7\%}$ & 1.64 & 1.75 & 4.54 \\
\hline
6 & 0.83 &  5.68 &  $\frac{1}{3}$ &  -0.5 &  $\frac{1}{3}$ & $179.4^{+20\%}_{-16\%}$ & $200.1^{+5.9\%}_{-9.3\%}$ & 2.16 & 2.41 & 5.45 \\
\hline
7 & 0.94 & -0.10 &  1& $\frac{1}{6}$ &  -$\frac{1}{6}$ &  $131.1^{+22\%}_{-17\%}$ & $146.2^{+12\%}_{-11\%}$ & 2.26 & 2.54 & 3.98 \\
\hline
\end{tabular}
}
\caption{
NLO and \nnloapp{} results for the benchmark points derived in Ref.~\cite{Capozi:2019xsi}. The values for the cross sections are given at $\sqrt{s}=14$\,TeV. The given uncertainties are scale uncertainties based on 3-point scale variations, see text for details. $K_{\rm{NLO}}$ and $K_{\rm{NNLO}^\prime}$ denote the corresponding ratio to the LO cross section. The ratio to the SM prediction in the last column is computed using the \nnloapp{} result.
}
\label{tab:benchmarks}
\end{table}

The \nnloapp{} result is rescaled by a constant factor of 0.958 in order to match the NNLO$_{\rm FTapprox}$ total cross section of Ref.~\cite{Grazzini:2018bsd} when the SM limit is taken.\footnote{
In this way we account for the effect of the  top-quark mass in the double-real and real-virtual amplitudes in the SM case, as included in NNLO$_{\rm FTapprox}$.}
We note that all other results in this paper (apart from the invariant-mass distributions in Fig.~\ref{fig:mhh}), including the parametrisation presented in the following section, are insensitive to this additional factor, which cancels out in the ratio $\sigma/\sigma_{\rm SM}$.

From the results in Table~\ref{tab:benchmarks} we can observe that the size of the QCD corrections, described by the \nnloapp{} $K$-factor, have a sizeable dependence on the EFT parameters. In the particular case of the seven benchmark points under consideration, the $K$-factor can differ from the SM one by almost $40\%$.
For all the EFT points in Table~\ref{tab:benchmarks} we can observe a sizeable reduction in the scale uncertainties once the \nnloapp{} corrections are included.

\subsection{Differential results and heat maps}

\begin{figure}
\begin{center}
\includegraphics[width=0.47\textwidth]{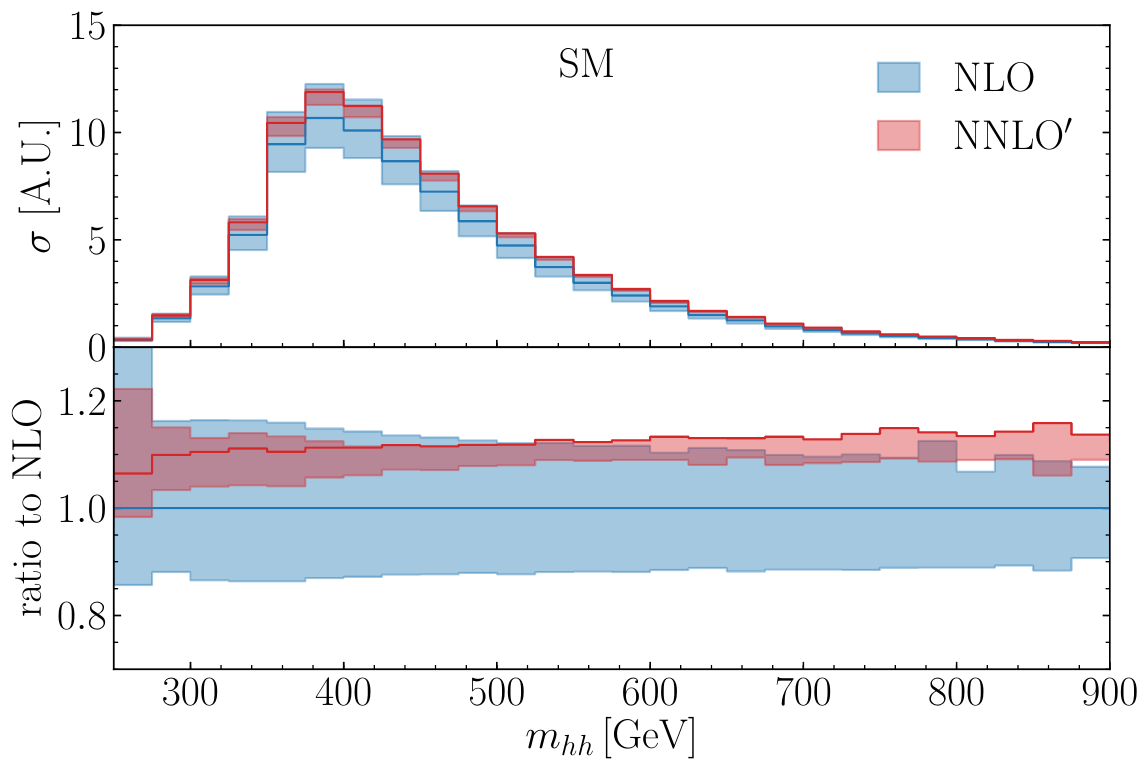}
\includegraphics[width=0.47\textwidth]{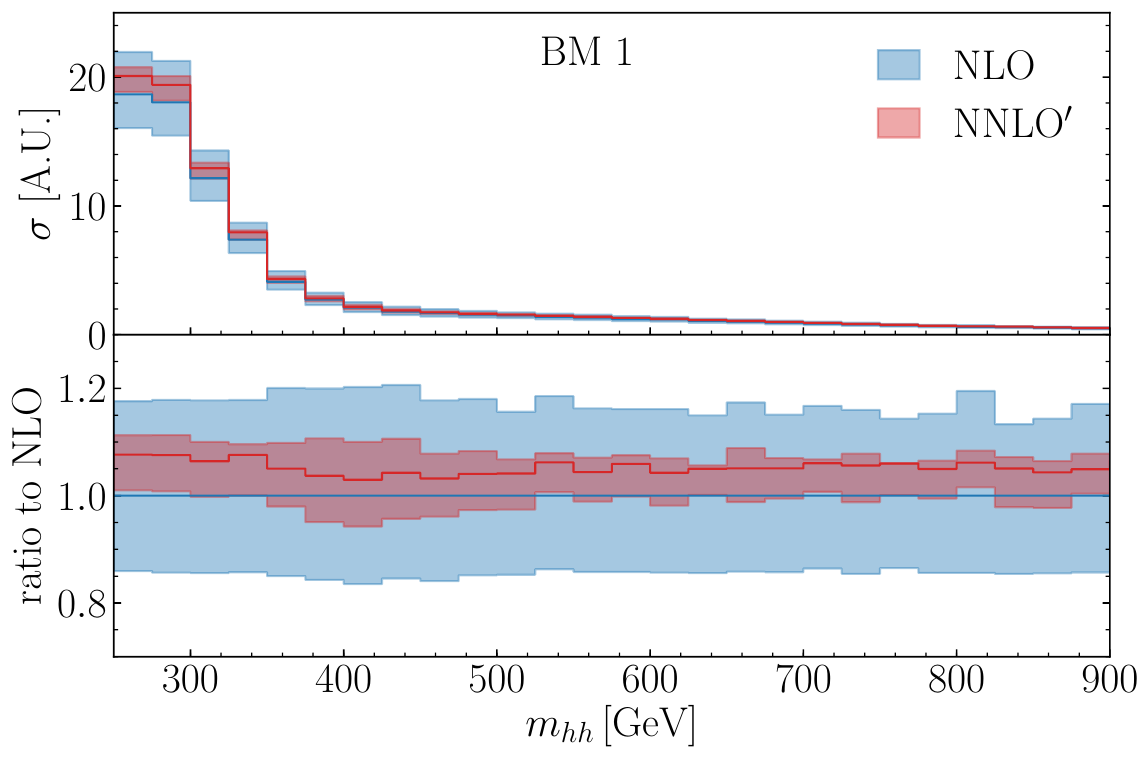}
\\
\includegraphics[width=0.47\textwidth]{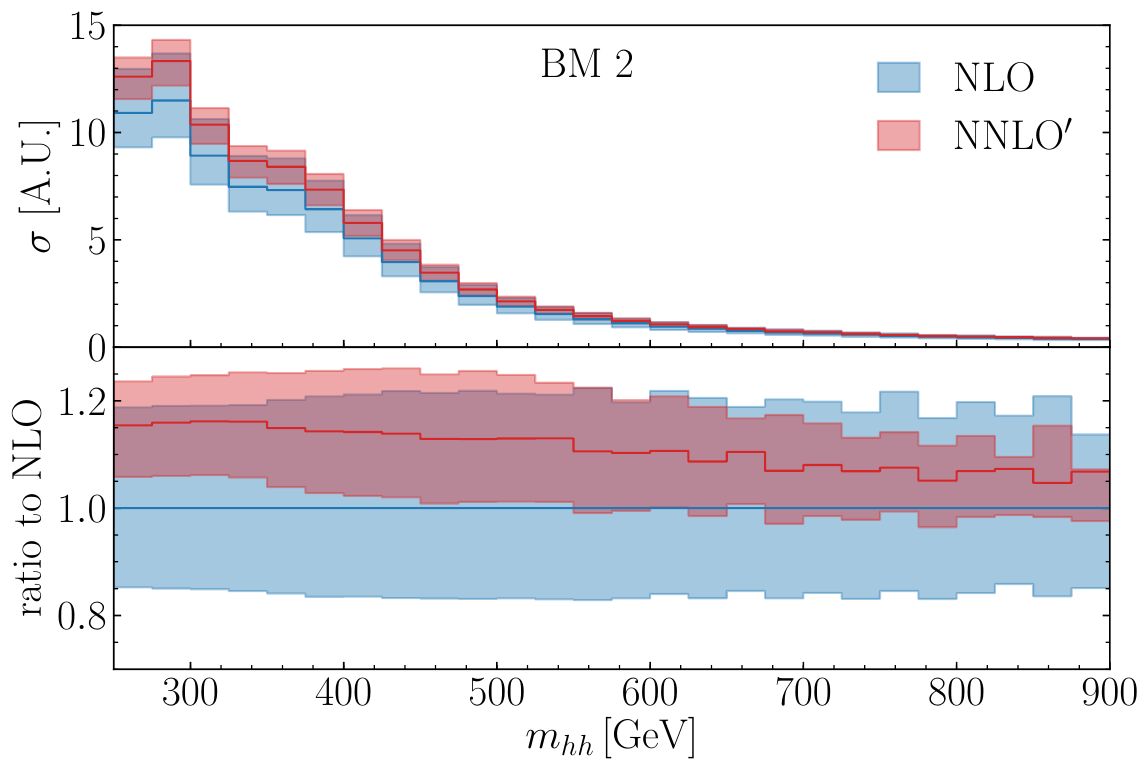}
\includegraphics[width=0.47\textwidth]{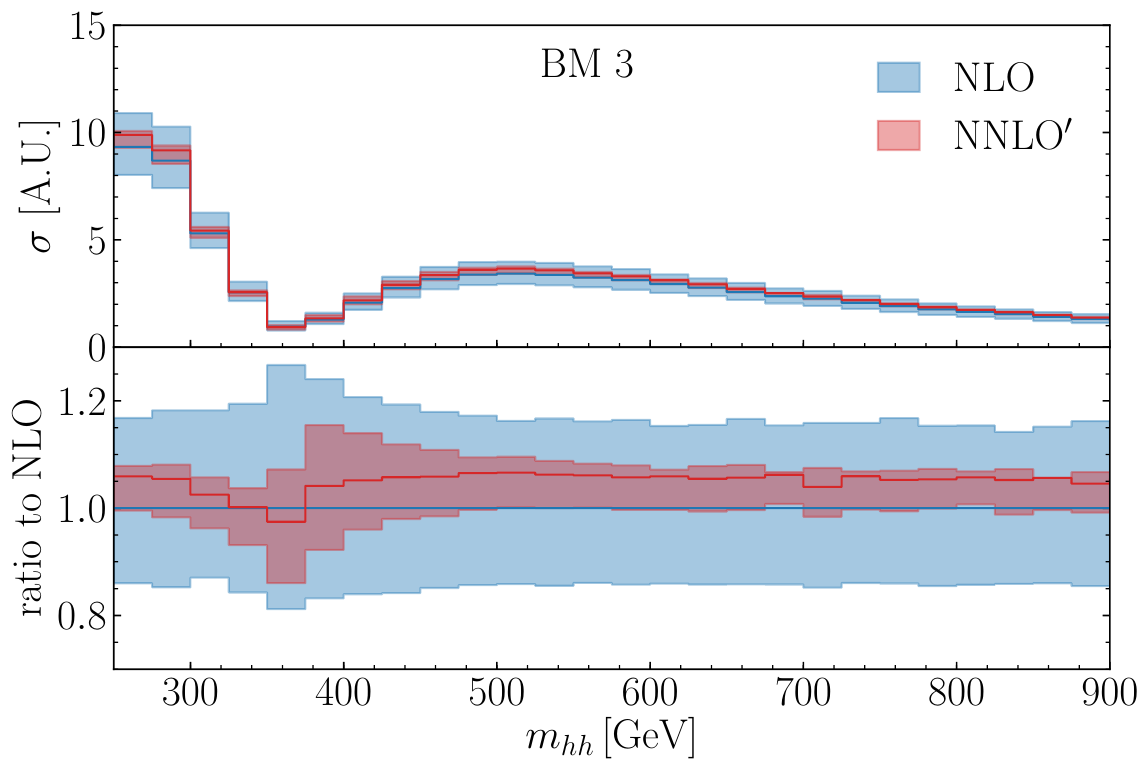}
\\
\includegraphics[width=0.47\textwidth]{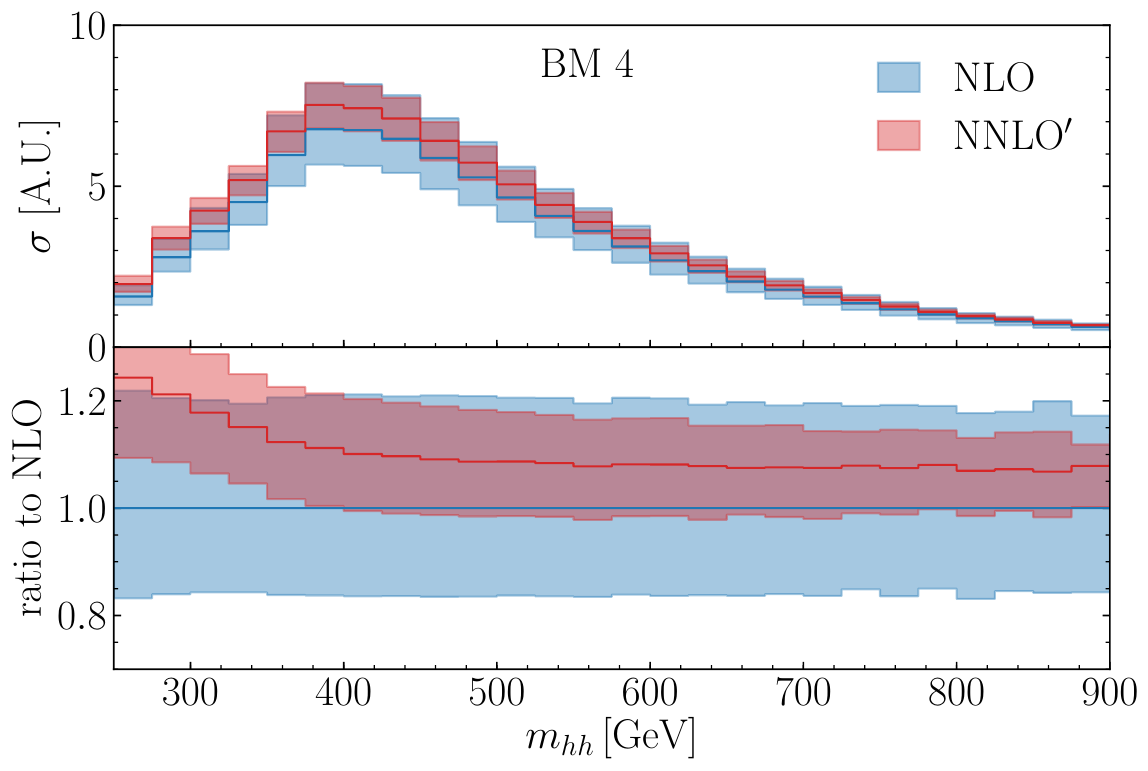}
\includegraphics[width=0.47\textwidth]{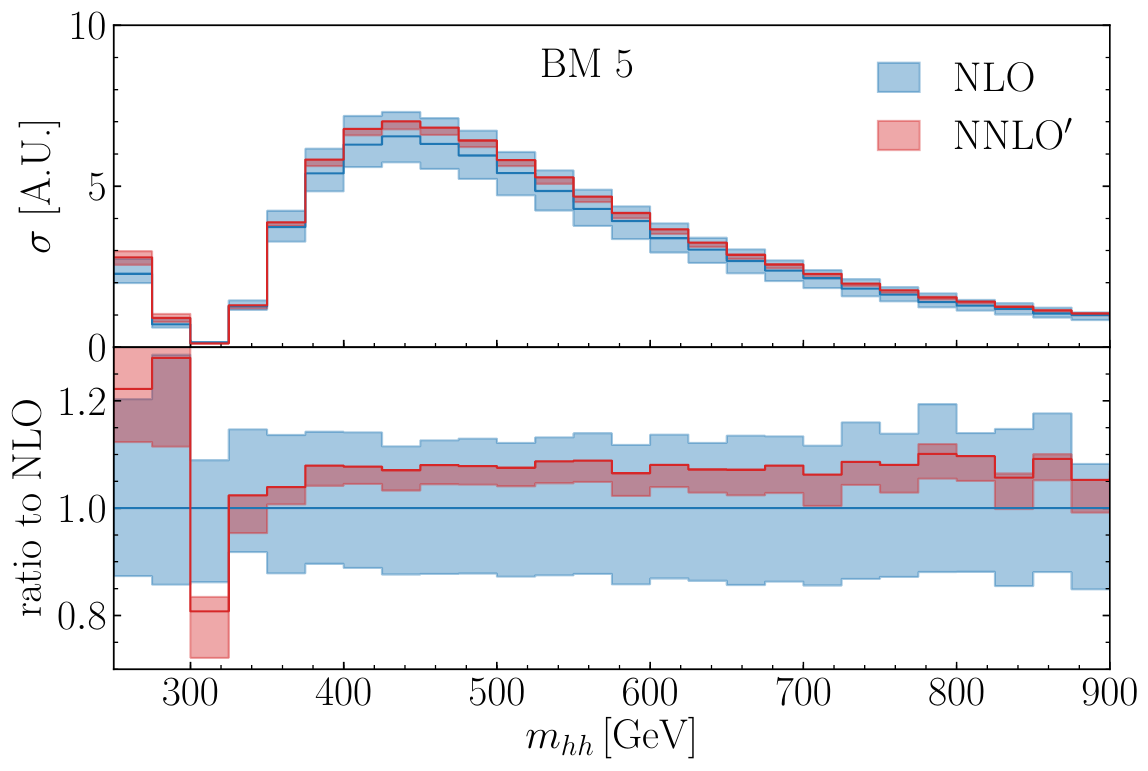}
\\
\includegraphics[width=0.47\textwidth]{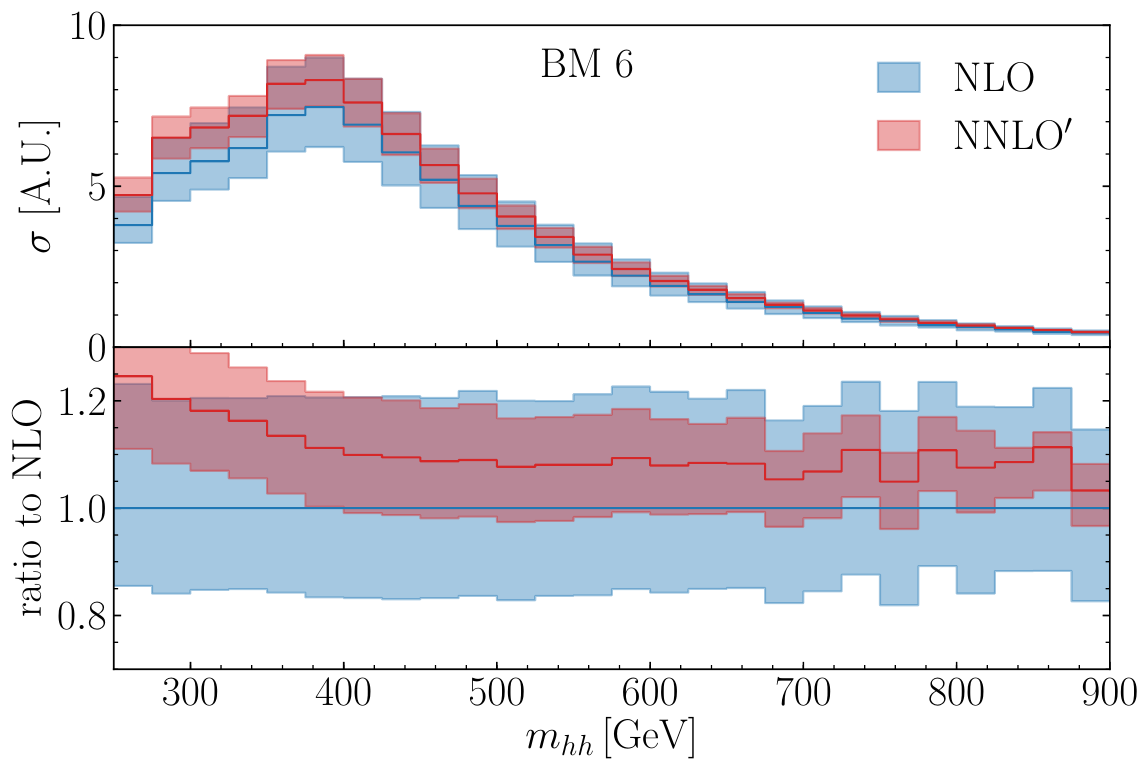}
\includegraphics[width=0.47\textwidth]{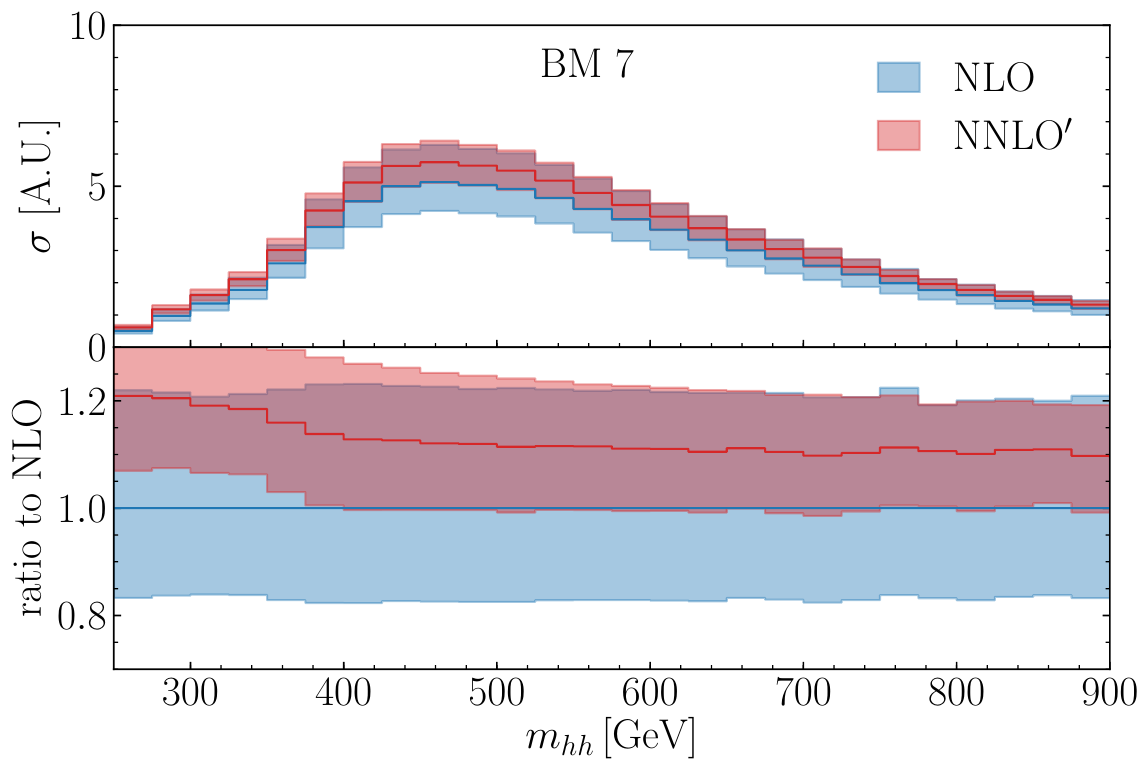}
\end{center}
\vspace{-0.5cm}
\caption{\label{fig:mhh}
Invariant-mass distribution of the di-Higgs system for the SM and the seven benchmark points in Table~\ref{tab:benchmarks}, at NLO (blue) and \nnloapp{} (red).
}
\end{figure}

In Fig.~\ref{fig:mhh} we present the di-Higgs invariant-mass distribution for the SM and for the seven shape benchmarks listed in Table~\ref{tab:benchmarks}, both at NLO and \nnloapp.

In Fig.~\ref{fig:mhh} we can observe a clear reduction of the scale uncertainties between NLO and \nnloapp. Furthermore, some of the benchmark points show a rather non-uniform differential \nnloapp/NLO $K$-factor, where in particular the region close to the Higgs-boson pair production threshold is enhanced compared to NLO, while in the SM case the enhancement is stronger in the tail of the distribution.
Note that the $\chhh$ value for both benchmark 3 and 4 is close to the value  $\chhh \simeq 2.5$ where the cross section goes through a minimum as a function of $\chhh$ only, and shows a dip in the $\mhh$ distribution if the other couplings are SM-like.
However, for benchmark 4, this dip is not present because the values of the other couplings destroy the strong cancellations that lead to the dip.

%%====================================
%%====================================

The low value of $\ct$ for benchmark 4 has the effect of reducing the contribution of the box-type diagrams to the cross section, such that the triangle-type contributions, dominating at low $\mhh$, play a larger role, even more so at \nnloapp{} due to the dominance of the soft-virtual contributions in this kinematic region. Note also that the coefficients $a_{24}$ and $a_{25}$ in Eq.~(\ref{eq:AcoeffsNNLO}), which are only present at NNLO, contain $\cg$ to the powers 4 and 3, respectively, therefore the influence of this coupling has a different functional dependence at NNLO, which can lead to shape distortions.

\begin{figure}[tb]
\begin{center}
\includegraphics[width=0.49\textwidth]{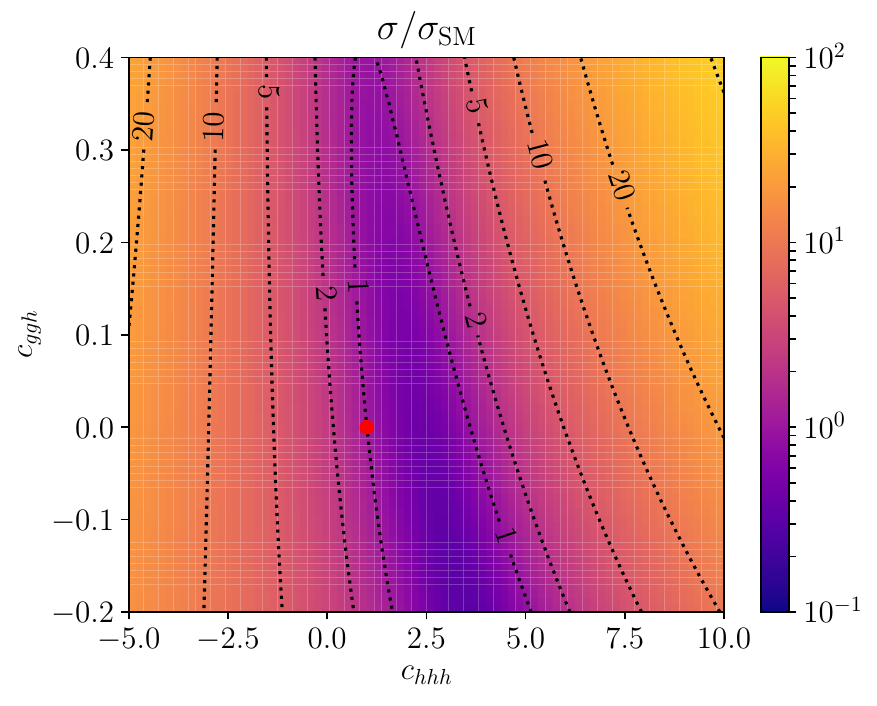}
\includegraphics[width=0.49\textwidth]{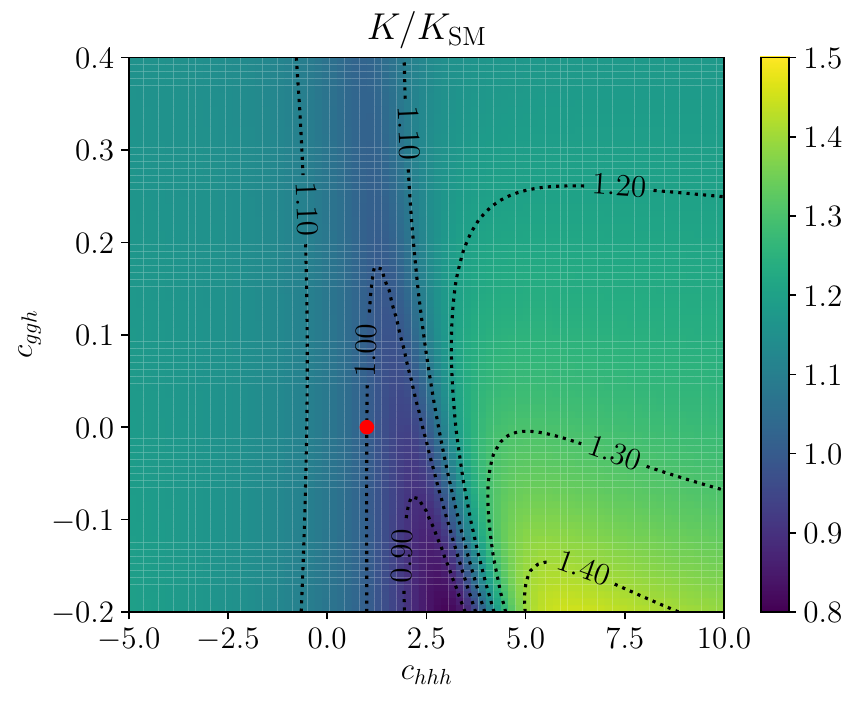}
\end{center}
\caption{\label{fig:cg_vs_c3_LO}
Ratio of (left) the total cross section to the SM cross section at \nnloapp{} and (right) the \nnloapp{}/LO $K$-factor to the SM $K$-factor, in the $\chhh-\cg$ plane. The red dot indicates the SM value for these couplings.
}
\end{figure}

\begin{figure}[tb]
\begin{center}
\includegraphics[width=0.49\textwidth]{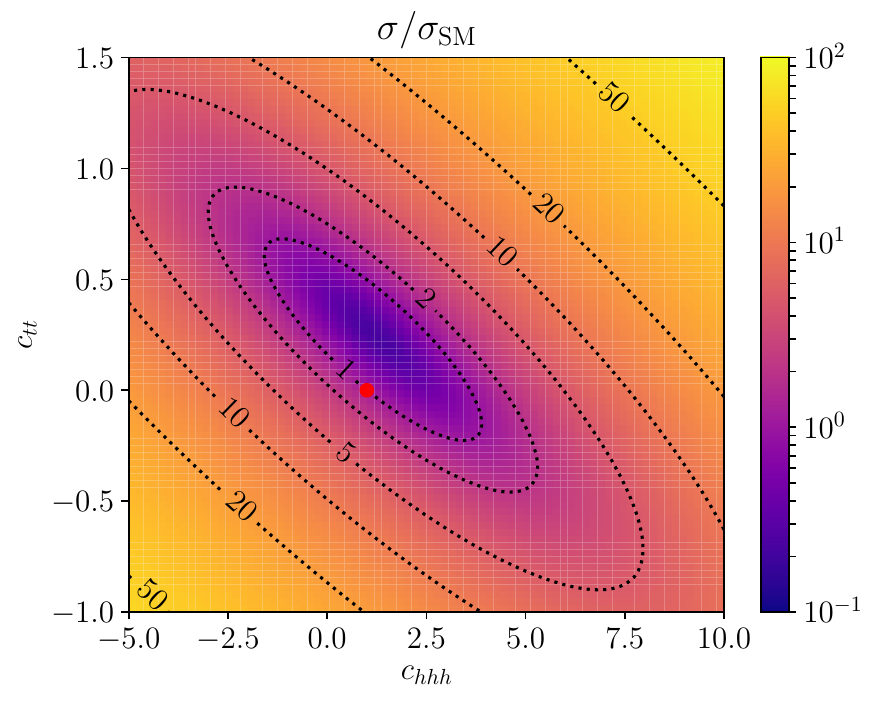}
\includegraphics[width=0.49\textwidth]{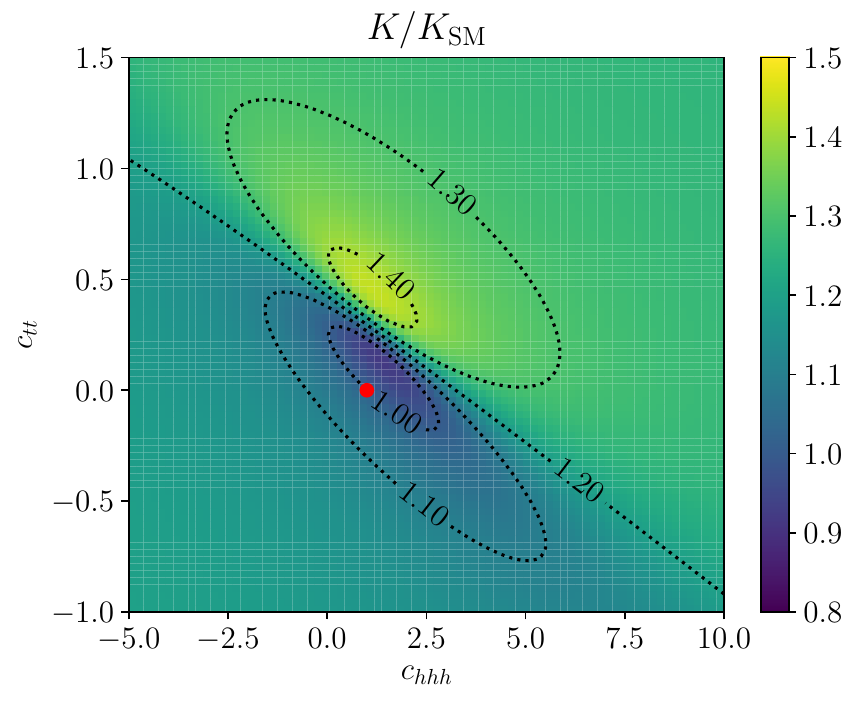}
\end{center}
\caption{\label{fig:ctt_vs_c3_LO}
Same as Fig.~\ref{fig:cg_vs_c3_LO} for the $\chhh-\ctt$ plane.
}
\end{figure}

In Fig.~\ref{fig:cg_vs_c3_LO}~(Fig.~\ref{fig:ctt_vs_c3_LO}) we present the \nnloapp{} total cross section and $K$-factor, both normalised to the corresponding SM value, in the $\chhh$--$\cg$ ($\chhh$--$\ctt$) plane.\footnote{We note that these plots have been produced based on the fit of the $a_i$ coefficients in Eq.~(\ref{eq:AcoeffsNNLO}) that is described in the following section.}
Here the $K$-factor is defined as the ratio between the \nnloapp{} and LO predictions.
The couplings are varied in the range $\chhh \in [-5,10]$, $\cg \in [-0.2,0.4]$ and $\ctt \in [-1,1.5]$, motivated by current limits~\cite{Sirunyan:2020xok,Aad:2020mkp,Sirunyan:2020icl,ATLAS-CONF-2021-016,Sirunyan:2021ybb,Sirunyan:2021fpv} in a conservative way.
In both cases we set $c_t = 1$ and the remaining couplings to zero, except in the $\chhh$--$\cg$ case where we additionally set $\cgg = \cg/2$ in order to mimic the SMEFT situation, where the latter couplings are not independent of each other.

Comparing Figs.~\ref{fig:cg_vs_c3_LO} and \ref{fig:ctt_vs_c3_LO} (left), we observe that the cross section is more sensitive to variations of both $\chhh$ and $\ctt$ than variations of $\cg$ (within the range suggested by current constraints). Related to this, we observe strong correlations of the cross-section values under simultaneous variations of these two parameters.
It is also clear that the normalised $K$-factors, shown in  Figs.~\ref{fig:cg_vs_c3_LO} and \ref{fig:ctt_vs_c3_LO} (right), only turn out to be rather flat, i.e. very similar to the SM ones, under variations of the parameter $\cg$, which generates point-like ($m_t$-independent) effective interactions. In contrast, much stronger $K$-factor changes (up to 40\%) are observed for modifications of $\chhh$ and $\ctt$, due to the effect of the full top-quark mass dependence in the NLO corrections. We point out that the $K$-factors shown here mostly have values below two because they are normalised to the SM $K$-factor.

%%====================================
%%====================================

%%====================================
%%====================================

\begin{figure}[tb]
\begin{center}
\includegraphics[width=0.49\textwidth]{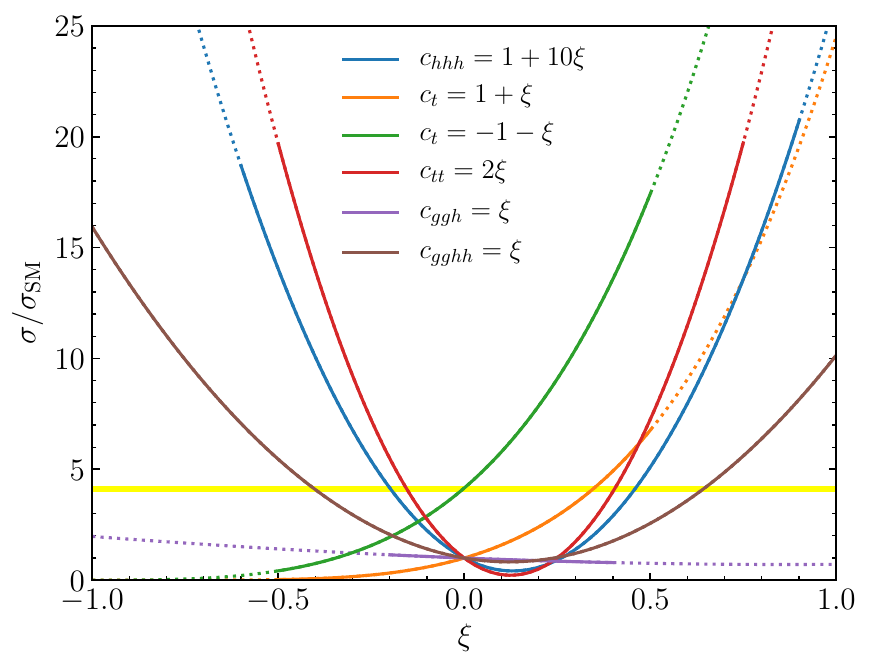}
\includegraphics[width=0.502\textwidth]{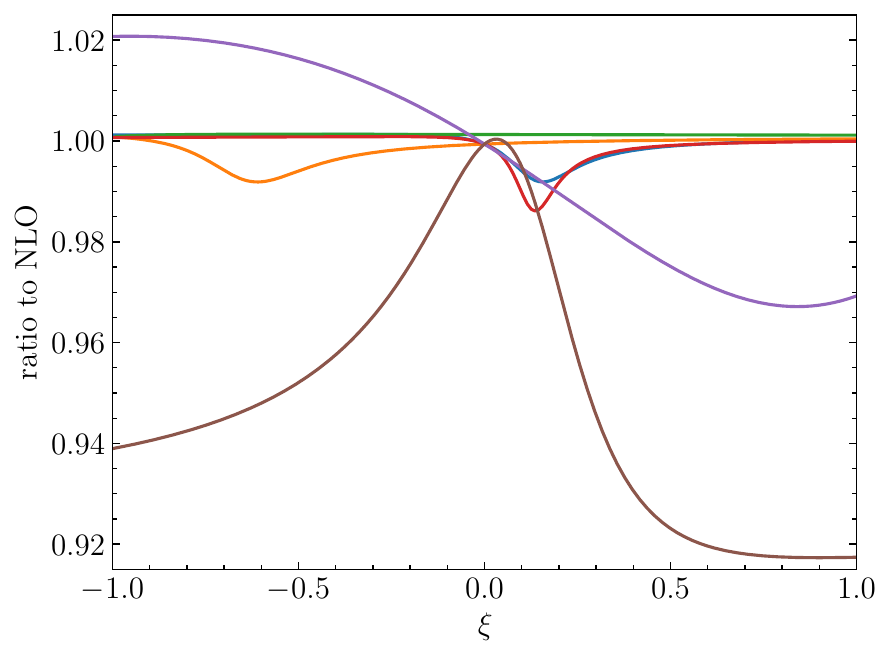}
\end{center}
\vspace{-0.5cm}
\caption{\label{fig:xs_vs_couplings}
Total cross section, normalised to its SM value, as a function of each of the anomalous couplings. The horizontal line corresponds to the best current experimental limit \cite{ATLAS-CONF-2021-016}.
The panel on the right shows the ratio to the NLO curves, that is $(\sigma/\sigma_{\rm SM})_{{\rm NNLO}'}/(\sigma/\sigma_{\rm SM})_{\rm NLO}$.
}
\end{figure}

In Fig.~\ref{fig:xs_vs_couplings} (left) we present the cross section as a function of each of the couplings present in the Lagrangian of Eq.~(\ref{eq:ewchl}).
We vary each of the couplings independently while keeping the others at their SM value.
The regions where the lines in Fig.~\ref{fig:xs_vs_couplings} (left) are solid correspond to the following range in the EFT parameter space:
\begin{equation}\label{eq:range}
    c_{hhh} \in [-5,10],\;
    |c_{t}| \in [0.5,1.5],\;
    c_{tt} \in [-1,1.5],\;
    c_{ggh} \in [-0.2,0.4],\;
    c_{gghh} \in [-1,1].
\end{equation}
The chosen range for each coefficient is motivated by current experimental constraints~\cite{ATLAS-CONF-2021-016,Sirunyan:2020xok,Aad:2020mkp,Sirunyan:2020icl,Sirunyan:2021ybb,Sirunyan:2021fpv}. However, since these limits are not based on a simultaneous fit of all the couplings under consideration here, we have conservatively increased the allowed range of variation.\footnote{The range for $\cg$ in Table~\ref{tab:benchmarks} is larger because at the time when the benchmark points were constructed the tighter constraints were not yet available.}
We note that in the case of $c_{gghh}$ the interval in Eq.~(\ref{eq:range}) is chosen arbitrarily, since there are no experimental constraints on its value.

%%====================================
%%====================================

We can compare the curves in Fig.~\ref{fig:xs_vs_couplings} (left) with the best experimental limit on the total di-Higgs production cross section of 4.1 times its SM value at 95\% confidence level~\cite{ATLAS-CONF-2021-016}, indicated by the yellow horizontal line.
From this comparison we conclude that, with the only exception of $c_{ggh}$, all of the anomalous couplings can, within the limits given in Eq.~(\ref{eq:range}), generate variations in the di-Higgs cross section which are larger than the current experimental limit.
While Eq.~(\ref{eq:range}) describes the allowed region in the EFT parameter space only in a qualitative way, the results in Fig.~\ref{fig:xs_vs_couplings} (left) clearly indicate that, with the present level of precision achieved by the LHC, a simultaneous variation of all couplings is needed for a meaningful EFT analysis.

The right panel of Fig.~\ref{fig:xs_vs_couplings} shows the ratio of the \nnloapp{} and NLO results for the same coupling modifications that are present in the left panel of the figure.
As we are comparing perturbative predictions for the ratio $\sigma/\sigma_{\rm SM}$, we can expect a reduced impact from higher-order corrections, since their effect will partially cancel between numerator and denominator (in particular, the quantity presented in the right panel of Fig.~\ref{fig:xs_vs_couplings} is, by definition, equal to 1 in the SM limit). In fact, we can observe that for some of the curves displayed in Fig.~\ref{fig:xs_vs_couplings} the difference between \nnloapp{} and NLO is below 1\%. Other coupling variations ($c_{ggh}$ and $c_{gghh}$) feature a larger deviation from the NLO prediction, the difference going up to 8\% in the case of $c_{gghh}$ for the considered range. We note that larger deviations between the \nnloapp{} and NLO predictions for $\sigma/\sigma_{\rm SM}$ can be obtained when all couplings are allowed to vary simultaneously.

%\clearpage

\subsection{Fitting procedure and results for the coupling coefficients}
\label{sec:fit}

In order to obtain a parametrisation of the \nnloapp{} total cross section as a function of the anomalous couplings, we perform a non-linear fit of the $a_i$ coefficients in Eq.~(\ref{eq:AcoeffsNNLO}).
The fit is carried out with the software {\sc Mathematica}, and is based on the values of the \nnloapp{} cross section computed for 43 different points in the EFT parameter space. The list of the points is shown in Appendix~\ref{sec:appendix}.  
We repeat the fit for the three scale choices of $\mu_R$ and $\mu_F$ described in the previous section. The result of the fits is shown in Table~\ref{tab:NNLOfitAllScales}. 

\begin{table}
\begin{center}
\begin{tabular}{|c|c|c|c|}
\hline
 & $\mu_R = \mu_F = \mu_0$ & $\mu_R = \mu_F = \mu_0/2$ & $\mu_R = \mu_F = 2\mu_0$ \\ \hline
$a_{1}$	&	2.2359	&	2.2899	&	2.1062	\\ \hline
$a_{2}$	&	12.465	&	13.191	&	11.469	\\ \hline
$a_{3}$	&	0.34254	&	0.36853	&	0.31341	\\ \hline
$a_{4}$	&	0.32832	&	0.27278	&	0.3225	\\ \hline
$a_{5}$	&	12.035	&	12.139	&	11.435	\\ \hline
$a_{6}$	&	-9.6736	&	-9.967	&	-9.0278	\\ \hline
$a_{7}$	&	-1.5785	&	-1.6626	&	-1.4625	\\ \hline
$a_{8}$	&	3.4554	&	3.621	&	3.2097	\\ \hline
$a_{9}$	&	2.8013	&	2.5608	&	2.6905	\\ \hline
$a_{10}$	&	16.173	&	16.712	&	15.144	\\ \hline
$a_{11}$	&	-1.1806	&	-1.2201	&	-1.0647	\\ \hline
$a_{12}$	&	-5.6581	&	-5.6718	&	-5.3808	\\ \hline
$a_{13}$	&	0.63134	&	0.65511	&	0.59109	\\ \hline
$a_{14}$	&	2.7664	&	2.9025	&	2.581	\\ \hline
$a_{15}$	&	2.93	&	2.9659	&	2.7499	\\ \hline
$a_{16}$	&	-0.10785	&	-0.14072	&	-0.12683	\\ \hline
$a_{17}$	&	0.223	&	0.52954	&	0.098154	\\ \hline
$a_{18}$	&	0.065656	&	0.032461	&	0.082079	\\ \hline
$a_{19}$	&	0.18294	&	0.22852	&	0.1622	\\ \hline
$a_{20}$	&	-0.048533	&	-0.056875	&	-0.02693	\\ \hline
$a_{21}$	&	0.12436	&	0.33443	&	0.036752	\\ \hline
$a_{22}$	&	0.027999	&	0.03496	&	0.022263	\\ \hline
$a_{23}$	&	0.21161	&	0.21764	&	0.15791	\\ \hline
$a_{24}$	&	0.00047051	&	0.00073051	&	0.00031311	\\ \hline
$a_{25}$	&	0.00077149	&	0.00087966	&	-0.00040697	\\ \hline
\end{tabular}
\end{center}
\caption{Fit results for the coefficients of the 25 coupling combinations in Eq.~(\ref{eq:AcoeffsNNLO}) at three different scale choices, with $\mu_0=m_{hh}/2$,
at a centre-of-mass energy of $\sqrt{s}=14$ TeV.
}
\label{tab:NNLOfitAllScales}
\end{table}

 We perform an inverse-variance weighted fit, using the uncertainty of the \nnloapp{} cross sections. This uncertainty is dominated by the statistical uncertainty associated with the determination of the top-mass-dependent two-loop virtual corrections, which are calculated numerically from a finite number of 6715 phase-space points, while the numerical uncertainties from the Monte Carlo integration are considerably smaller. 
 Because the phase-space points were originally generated to populate the $m_{hh}$ distribution for SM values of the couplings, the phase-space cover is suboptimal in regions where the differential cross section is much larger than in the SM case.
In order to estimate this uncertainty, we propagate the statistical uncertainty stemming from the virtual contribution to the total differential cross section 
  in $m_{hh}$, for all of the 43 EFT points used in the fit, see Tables~\ref{tab:fitPoints1} and~\ref{tab:fitPoints2}. Further details about the choice of the EFT points are given in Appendix \ref{sec:appendix}.

This uncertainty of the NLO contribution is $\leq 2\%$ in all bins, except in the first $m_{hh}$ bin. 
There it can reach $\sim 20\%$ for a few points, and is otherwise around $4-12\%$ for most points.
Thus the uncertainty on the total cross section is typically at the percent level for the majority of the points used in the fit.
If larger uncertainties are found for some points in the EFT parameter space they feature either:
\begin{itemize}
    \item large numerical cancellations between different contributions at NLO,
    \item an enhancement of the low-$m_{hh}$ region (and to a lesser extent of the tail for benchmark points with non-zero values of $\cg$ and $\cgg$). These regions, being suppressed in the SM case, are less densely populated by the numerical grid used to compute the NLO virtual corrections,
    \item or a proportionally large contribution from the virtual corrections to the total cross section.
\end{itemize}

In order to check the robustness of our fit and to estimate the uncertainties associated to it, we produce 1000 replicas of the 43 cross section values entering the fit, by randomly generating Gaussian variations around their central value, and using the corresponding uncertainty as the standard deviation. Based on these replicas, we in turn obtain 1000 replicas of the fitted cross section.
Finally, we perform a scan in the EFT parameter space, randomly generating 10000 points in the range indicated in Eq.~(\ref{eq:range}).

We study the variation in the cross section by computing its standard deviation over the 1000 fit replicas, for each of the 10000 points in the EFT parameter scan. We note that this standard deviation can be regarded as an estimate of the final uncertainty in the cross section, encompassing both the uncertainties in the 43 original values of the cross section as well as the uncertainties coming from the fitting procedure. We have found these standard deviations to be at the percent level. Specifically, for $68\%$ ($95\%$) of the points in the EFT scan this deviation is below $2.0\%$ ($5.8\%$).

Based on this analysis, we conclude that the uncertainties in the cross sections that are obtained by using Eq.~(\ref{eq:AcoeffsNNLO}) together with the results in Table~\ref{tab:NNLOfitAllScales} are at the ${\cal O}(5\%)$ level. We note that this uncertainty is of the same order of magnitude as the one in the cross section values used to perform the fit. In other words, the fitting procedure does not produce a substantial increase in the uncertainty.

A more detailed estimation of the numerical uncertainties is beyond the scope of this paper. We note, however, that our ${\cal O}(5\%)$ estimation is mostly below the scale uncertainties at \nnloapp, as can be seen from Table~\ref{tab:benchmarks}, and also below uncertainties related to the renormalisation scheme dependence of the top-quark mass, which are estimated to be between $+4\%$ and $-18\%$ at total cross section level in the SM case~\cite{Baglio:2020wgt}.

%\clearpage

\section{Conclusions}
\label{sec:conclusions}
We have presented a combination of NLO predictions with full top-quark mass dependence with approximate NNLO corrections for Higgs-boson pair production in gluon fusion. Anomalous couplings are included in the framework of an Effective Field Theory where the dominant operators contributing to this process are parametrised by five anomalous couplings, $\chhh, \ct, \ctt, \cg$ and $\cgg$.
Our combination, which we call \nnloapp, is based on an additive approach of the full NLO corrections and the NNLO corrections in the heavy top limit improved by inserting LO form factors with full top-quark mass dependence.

We have given \nnloapp{} results including scale uncertainties at $\sqrt{s}=14$\,TeV at seven benchmark points, for the total cross section as well as for the $\mhh$ distribution.
We observe a strong reduction of the scale uncertainties compared to the NLO case, by a factor of 2--3, depending on the benchmark point. The differential results also show shape distortions at \nnloapp{} for some of the benchmark points, in particular close to the Higgs-boson pair production threshold.

Parametrising the cross section in terms of all possible coupling combinations leads to 25 coupling coefficients $a_i$, of which two combinations occur at NNLO which are not present at NLO (while 10 new combinations occur comparing LO and NNLO). 
Based on the above parametrisation of the cross section in terms of coupling combinations, we have produced heat maps for slices of the coupling-parameter space, for both the cross section and the $K$-factor (normalised to their SM values), where we varied the anomalous couplings in a range motivated by current constraints.
The feature that the cross section is much more sensitive to variations of $\chhh$ and $\ctt$ than of $\cg$, found already at NLO, also can be seen at \nnloapp.
Furthermore, we presented individual variations of couplings and compared them to the current best limit on the total cross section, which leads to the conclusion that simultaneous variations of all couplings are needed for a meaningful EFT analysis.
We also provided the values for the $a_i$ coefficients at three different scale choices, thereby allowing fast and flexible studies of anomalous coupling variations at \nnloapp{} level.

\section*{Acknowledgements}
We would like to thank Stephen Jones, Matthias Kerner and Johannes
Schlenk for collaboration on the NLO {\tt ggHH} results and code.
This research was supported in part by the COST Action CA16201 (`Particleface') of the European Union
and  by  the  Deutsche  Forschungsgemeinschaft (DFG, German Research
Foundation) under grant 396021762 - TRR 257. LS is supported by the Royal Society under 
grant number RP\textbackslash R1\textbackslash 180112 and by Somerville College.
DdeF and IF are partially supported by CONICET and ANPCyT.
We also acknow\-ledge resources provided by the Max Planck Computing and Data Facility (MPCDF). 

%\newpage                              
\appendix
\section{Appendix}                                                               
\label{sec:appendix} 

We summarise in Tables~\ref{tab:fitPoints1} and~\ref{tab:fitPoints2} the EFT points that enter 
the fit described in Section~\ref{sec:fit}. These points are associated with the following
different publications:
\begin{itemize}
    \item Points 1--23 (Table~\ref{tab:fitPoints1}): used in Ref.~\cite{Heinrich:2020ckp} as 
    an interpolation basis for the virtual grids at NLO, which were interfaced to the \POWHEGBOX,
    \item Points 24--30 (Table~\ref{tab:fitPoints1}): the seven $m_{hh}$-characteristic benchmark 
    points identified in Ref.~\cite{Capozi:2019xsi} at NLO and appearing in Table~\ref{tab:benchmarks},
    \item Points 31--36 (Table~\ref{tab:fitPoints2}): pure $\chhh$ variations computed at NLO in 
    Ref.~\cite{Heinrich:2019bkc},
    \item Points 37--43 (Table~\ref{tab:fitPoints2}): additional points which were computed to 
    help decrease the statistical uncertainty on the fit of the cross section, in particular 
    the one associated with the (small) NNLO $a_i$ coefficients.
\end{itemize}

\begin{table}[!ht]
\renewcommand{\arraystretch}{1.1}
\begin{center}
\begin{tabular}{|c|c|c|c|c|c|}
\hline
 \#  & $c_{t}$ & $c_{hhh}$ & $c_{tt}$ & $c_{ggh}$ & $c_{gghh}$ \\
\hline
1	&	$\frac{11}{23}$	&	$\frac{1}{2}$	&	$1$	&	$\frac{11}{16}$	&	$\frac{8}{9}$	\\ \hline
2	&	$-\frac{5}{2}$	&	$-1$	&	$\frac{2}{7}$	&	$\frac{7}{15}$	&	$\frac{2}{11}$	\\ \hline
3	&	$-\frac{1}{19}$	&	$\frac{1}{22}$	&	$-\frac{11}{16}$	&	$\frac{5}{17}$	&	$\frac{9}{13}$	\\ \hline
4	&	$\frac{1}{9}$	&	$\frac{1}{8}$	&	$-\frac{1}{3}$	&	$\frac{5}{23}$	&	$-\frac{4}{5}$	\\ \hline
5	&	$\frac{4}{9}$	&	$-\frac{8}{7}$	&	$\frac{1}{11}$	&	$\frac{8}{11}$	&	$\frac{9}{14}$	\\ \hline
6	&	$\frac{5}{11}$	&	$\frac{5}{6}$	&	$-\frac{3}{7}$	&	$\frac{2}{21}$	&	$\frac{10}{19}$	\\ \hline
7	&	$\frac{1}{7}$	&	$1$	&	$-\frac{1}{3}$	&	$\frac{12}{23}$	&	$-\frac{3}{2}$	\\ \hline
8	&	$-\frac{1}{14}$	&	$-\frac{1}{19}$	&	$\frac{5}{24}$	&	$\frac{9}{13}$	&	$\frac{9}{10}$	\\ \hline
9	&	$\frac{8}{17}$	&	$-\frac{3}{8}$	&	$-\frac{3}{8}$	&	$-\frac{2}{3}$	&	$\frac{5}{12}$	\\ \hline
10	&	$-\frac{5}{6}$	&	$-\frac{11}{17}$	&	$\frac{1}{2}$	&	$-\frac{4}{3}$	&	$\frac{1}{2}$	\\ \hline
11	&	$\frac{1}{6}$	&	$-\frac{2}{3}$	&	$\frac{11}{10}$	&	$\frac{1}{2}$	&	$1$	\\ \hline
12	&	$\frac{3}{5}$	&	$\frac{1}{5}$	&	$\frac{2}{23}$	&	$-\frac{8}{23}$	&	$-\frac{4}{3}$	\\ \hline
13	&	$\frac{4}{9}$	&	$-\frac{8}{19}$	&	$-\frac{2}{9}$	&	$\frac{8}{21}$	&	$-\frac{4}{7}$	\\ \hline
14	&	$\frac{11}{5}$	&	$-\frac{2}{17}$	&	$-\frac{1}{13}$	&	$-\frac{3}{16}$	&	$\frac{5}{9}$	\\ \hline
15	&	$-\frac{6}{7}$	&	$\frac{1}{5}$	&	$-1$	&	$\frac{5}{16}$	&	$-\frac{7}{6}$	\\ \hline
16	&	$\frac{1}{9}$	&	$\frac{3}{10}$	&	$\frac{9}{7}$	&	$\frac{9}{7}$	&	$-\frac{6}{13}$	\\ \hline
17	&	$-\frac{4}{5}$	&	$-\frac{10}{23}$	&	$\frac{1}{9}$	&	$-\frac{12}{19}$	&	$\frac{1}{23}$	\\ \hline
18	&	$-\frac{1}{3}$	&	$-\frac{8}{7}$	&	$-\frac{1}{2}$	&	$-\frac{1}{2}$	&	$\frac{7}{17}$	\\ \hline
19	&	$-\frac{2}{3}$	&	$\frac{9}{4}$	&	$\frac{3}{11}$	&	$\frac{5}{14}$	&	$-1$	\\ \hline
20	&	$\frac{5}{18}$	&	$\frac{11}{18}$	&	$\frac{1}{9}$	&	$-\frac{4}{5}$	&	$\frac{5}{22}$	\\ \hline
21	&	$3$	&	$\frac{5}{23}$	&	$-\frac{10}{19}$	&	$\frac{1}{21}$	&	$-\frac{8}{21}$	\\ \hline
22	&	$\frac{2}{5}$	&	$\frac{5}{11}$	&	$-\frac{3}{2}$	&	$\frac{6}{11}$	&	$\frac{9}{14}$	\\ \hline
23	&	$\frac{10}{9}$	&	$\frac{1}{4}$	&	$-\frac{5}{12}$	&	$-\frac{4}{9}$	&	$\frac{1}{20}$	\\ \hline\hline
24	&	$\frac{47}{50}$	&	$\frac{197}{50}$	&	$-\frac{1}{3}$	&	$\frac{1}{2}$	&	$\frac{1}{3}$	\\ \hline
25	&	$\frac{61}{100}$	&	$\frac{171}{25}$	&	$\frac{1}{3}$	&	$0$	&	$-\frac{1}{3}$	\\ \hline
26	&	$\frac{21}{20}$	&	$\frac{221}{100}$	&	$-\frac{1}{3}$	&	$\frac{1}{2}$	&	$\frac{1}{2}$	\\ \hline
27	&	$\frac{61}{100}$	&	$\frac{279}{100}$	&	$\frac{1}{3}$	&	$-\frac{1}{2}$	&	$\frac{1}{6}$	\\ \hline
28	&	$\frac{117}{100}$	&	$\frac{79}{20}$	&	$-\frac{1}{3}$	&	$\frac{1}{6}$	&	$-\frac{1}{2}$	\\ \hline
29	&	$\frac{83}{100}$	&	$\frac{142}{25}$	&	$\frac{1}{3}$	&	$-\frac{1}{2}$	&	$\frac{1}{3}$	\\ \hline
30	&	$\frac{47}{50}$	&	$-\frac{1}{10}$	&	$1$	&	$\frac{1}{6}$	&	$-\frac{1}{6}$	\\ \hline
\end{tabular}
\end{center}
\caption{Points 1--30 in the EFT parameter space used for the fit.}
\label{tab:fitPoints1}
\end{table}

\begin{table}[!ht]
\renewcommand{\arraystretch}{1.1}
\begin{center}
\begin{tabular}{|c|c|c|c|c|c|}
\hline
 \#  & $c_{t}$ & $c_{hhh}$ & $c_{tt}$ & $c_{ggh}$ & $c_{gghh}$ \\
\hline
31	&	$1$	&	$-1$	&	$0$	&	$0$	&	$0$	\\ \hline
32	&	$1$	&	$0$	&	$0$	&	$0$	&	$0$	\\ \hline
33	&	$1$	&	$2$	&	$0$	&	$0$	&	$0$	\\ \hline
34	&	$1$	&	$\frac{12}{5}$	&	$0$	&	$0$	&	$0$	\\ \hline
35	&	$1$	&	$3$	&	$0$	&	$0$	&	$0$	\\ \hline
36	&	$1$	&	$5$	&	$0$	&	$0$	&	$0$	\\ \hline\hline
37	&	$0$	&	$1$	&	$0$	&	$5$	&	$0$	\\ \hline
38	&	$1$	&	$0$	&	$0$	&	$5$	&	$0$	\\ \hline
39	&	$0$	&	$0$	&	$0$	&	$1$	&	$0$	\\ \hline
40	&	$1$	&	$0$	&	$0$	&	$8$	&	$0$	\\ \hline
41	&	$-\frac{1}{2}$	&	$1$	&	$0$	&	$1$	&	$0$	\\ \hline
42	&	$1$	&	$1$	&	$0$	&	$1$	&	$0$	\\ \hline
43	&	$0$	&	$6$	&	$0$	&	$\frac{1}{10}$	&	$0$	\\ \hline
\end{tabular}
\end{center}
\caption{Points 31--43 in the EFT parameter space used for the fit.
}
\label{tab:fitPoints2}
\end{table}

\bibliographystyle{JHEP}

\bibliography{refs_HH_NNLOapprox}

\end{document}